\documentclass[a4paper]{article}

\usepackage{amsmath}
\usepackage{amssymb}
\usepackage{amsthm}
\usepackage{nccmath}	
\usepackage{mathrsfs}
\usepackage{wasysym}
\usepackage{bbm}		
\usepackage{enumitem}
\usepackage{mathtools}
\mathtoolsset{showonlyrefs=true}		
\usepackage{braket}		
\usepackage[normalem]{ulem}	
\usepackage{nicefrac}		
\usepackage{cancel}			
\usepackage{xcolor} 
\usepackage[a4paper, left=2.7cm, right=2.7cm, top=3.5cm, bottom=2cm]{geometry}
\usepackage[pdftex=true,hyperindex=true,pdfborder={0 0 0},colorlinks=true,allcolors=blue]{hyperref}
\usepackage{doi}
\usepackage[numbers]{natbib}

\usepackage{tikz}
\usetikzlibrary{shapes,arrows,automata,matrix,fit,patterns}
\usepackage{etoolbox}
\usepackage{calc}
\usepackage{xstring}

\newdimen\tilesize
\newdimen\tilecorner
\newcommand{\ctile}[2]{
	\begin{scope}
		\clip #2 +(-\tilesize,-\tilesize) rectangle +(\tilesize,\tilesize);%
		\IfEqCase{#1}{%
			{0}{%
				\fill[white] #2 +(-\tilesize,-\tilesize) rectangle +(\tilesize,\tilesize);%
				\draw[thin] #2 +(-\tilesize,-\tilesize) rectangle +(\tilesize,\tilesize);}%
			{1}{%
				\fill[black] #2 +(-\tilesize,-\tilesize) rectangle +(\tilesize,\tilesize);%
				\draw[thin] #2 +(-\tilesize,-\tilesize) rectangle +(\tilesize,\tilesize);
			}%
		}[%
			{%
			\fill[#1] #2 +(-\tilesize,-\tilesize) rectangle +(\tilesize,\tilesize);%
			\draw[thin] #2 +(-\tilesize,-\tilesize) rectangle +(\tilesize,\tilesize);%
			}%
		]
	\end{scope}
}

\newcommand{\colorededge}[3]{
	\begin{scope}[rounded corners=\tilecorner]
		\IfEqCase{#1}{%
			{l}{%
				\clip #3 -- +(-\tilesize,-\tilesize) -- +(-\tilesize,\tilesize) -- cycle;
				\fill[#2] #3 -- +(-\tilesize,-\tilesize) -- +(-\tilesize,\tilesize) -- cycle;
				\draw[very thin] #3 -- +(-\tilesize,-\tilesize) -- +(-\tilesize,\tilesize) -- cycle;
			}
			{r}{%
				\clip #3 -- +(\tilesize,-\tilesize) -- +(\tilesize,\tilesize) -- cycle;
				\fill[#2] #3 -- +(\tilesize,-\tilesize) -- +(\tilesize,\tilesize) -- cycle;
				\draw[very thin] #3 -- +(\tilesize,-\tilesize) -- +(\tilesize,\tilesize) -- cycle;
			}
			{d}{%
				\clip #3 -- +(-\tilesize,-\tilesize) -- +(\tilesize,-\tilesize) -- cycle;
				\fill[#2] #3 -- +(-\tilesize,-\tilesize) -- +(\tilesize,-\tilesize) -- cycle;
				\draw[very thin] #3 -- +(-\tilesize,-\tilesize) -- +(\tilesize,-\tilesize) -- cycle;
			}
			{u}{%
				\clip #3 -- +(-\tilesize,\tilesize) -- +(\tilesize,\tilesize) -- cycle;
				\fill[#2] #3 -- +(-\tilesize,\tilesize) -- +(\tilesize,\tilesize) -- cycle;
				\draw[very thin] #3 -- +(-\tilesize,\tilesize) -- +(\tilesize,\tilesize) -- cycle;
			}
		}[%
			\PackageError{colorededge}{Undefined tile side}{See the definition!}
		]
	\end{scope}
}

\newcommand{\wtile}[5]{
	\begin{scope}
		\colorededge{l}{#1}{#5}%
		\colorededge{r}{#2}{#5}%
		\colorededge{d}{#3}{#5}%
		\colorededge{u}{#4}{#5}%
	\end{scope}
}

\newcommand{\ama}{red}
\newcommand{\amb}{green}
\newcommand{\amc}{blue!30}
\newcommand{\amd}{yellow}
\newcommand{\ame}{gray}
\newcommand{\amf}{blue!70!red}

\newcommand{\amtile}[2]{
	\IfEqCase{#1}{%
		{1}{
			\wtile{\ama}{\amb}{\amb}{\ama}{#2}
		}
		{2}{%
			\wtile{\amc}{\amd}{\amd}{\amc}{#2}
		}
		{3}{%
			\wtile{\amd}{\ame}{\ame}{\amd}{#2}
		}
		{4}{%
			\wtile{\amf}{\amc}{\amc}{\amf}{#2}
		}
		{5}{%
			\wtile{\amc}{\ame}{\amd}{\amd}{#2}
		}
		{6}{%
			\wtile{\amc}{\amc}{\amd}{\amf}{#2}
		}
		{7}{%
			\wtile{\amd}{\amd}{\ame}{\amc}{#2}
		}
		{8}{%
			\wtile{\amf}{\amd}{\amc}{\amc}{#2}
		}
		{9}{%
			\wtile{\amb}{\ama}{\amc}{\ame}{#2}
		}
		{10}{%
			\wtile{\amb}{\ama}{\amf}{\amd}{#2}
		}
		{11}{%
			\wtile{\ama}{\ama}{\amd}{\ame}{#2}
		}
		{12}{%
			\wtile{\amb}{\amb}{\amf}{\amc}{#2}
		}
		{13}{%
			\wtile{\amd}{\amf}{\ama}{\amb}{#2}
		}
		{14}{%
			\wtile{\ame}{\amc}{\ama}{\amb}{#2}
		}
		{15}{%
			\wtile{\amc}{\amf}{\amb}{\amb}{#2}
		}
		{16}{%
			\wtile{\ame}{\amd}{\ama}{\ama}{#2}
		}
	}
}

\newdimen\blanksize
\newcommand{\xblank}[1]{
	\draw[very thin] #1 circle (\blanksize);
}

\theoremstyle{plain}
\newtheorem{theorem}{Theorem}[section]
\newtheorem{manualtheoreminner}{Theorem}
\newenvironment{manualtheorem}[1]{
  \renewcommand\themanualtheoreminner{#1}%
  \manualtheoreminner
}{\endmanualtheoreminner}
\newtheorem{lemma}[theorem]{Lemma}

\newtheorem{observation}[theorem]{Observation}

\theoremstyle{definition}

\makeatletter
\newenvironment{argumentenv}{%
	\list{}{%
		\leftmargin 4em%
		\listparindent 1em%
		\itemindent \listparindent
		\parsep        \z@ \@plus\p@}%
		\item\relax%
}{\endlist}
\makeatother

\newcommand{\ZZ}{\mathbb{Z}}			
\newcommand{\NN}{\mathbb{N}}			
\newcommand{\RR}{\mathbb{R}}			

\newcommand{\symb}[1]{\mathtt{#1}}		
\newcommand{\blank}{\diamond}			
\newcommand{\isdef}{\coloneqq}			
\DeclarePairedDelimiter\abs{\lvert}{\rvert}		

\newcommand{\ee}{\mathrm{e}}			
\newcommand{\PP}{\operatorname{\mathbb{P}}}		

\newcommand{\family}[1]{\mathscr{#1}}	

\newcommand{\unif}[1]{\underline{#1}}	
\newcommand{\unifO}{\unif{\symb{0}}}	

\newcommand{\majority}{
	\operatorname{majority}%
}

\begin{document} 

\title{
    Quasicrystalline Gibbs states in 4-dimensional lattice-gas models with finite-range interactions
} 
\author{
    Jacek Mi\c{e}kisz\thanks{
        University of Warsaw, Institute of Applied Mathematics and Mechanics, Banacha 2, 02-097 Warsaw, Poland.\\
        Email address: \texttt{\href{mailto:miekisz@mimuw.edu.pl}{miekisz@mimuw.edu.pl}}
    }
    \and
    Siamak Taati\thanks{
        Department of Mathematics \& Center for Advanced Mathematical Sciences, American University of Beirut, Beirut, Lebanon.
        Email address: \texttt{\href{mailto:siamak.taati@aub.edu.lb}{siamak.taati@aub.edu.lb}}
    }
} 
\date{}

\maketitle 

\begin{abstract}
We construct a four-dimensional lattice-gas model with finite-range interactions that has non-periodic, ``quasicrystalline'' Gibbs states at low temperatures. Such Gibbs states are probability measures which are small perturbations of non-periodic ground-state configurations corresponding to stacked tilings of the plane with Ammann's aperiodic tiles. Our construction is based on the correspondence between probabilistic cellular automata and Gibbs measures on their space-time trajectories, and a classical result on noise-resilient computing with cellular automata. The cellular automaton is constructed on the basis of Ammann's tiles, which are deterministic in one direction, and has non-periodic space-time trajectories corresponding to each valid tiling. Repetitions along two extra dimensions, together with an error-correction mechanism based on Toom's model, ensure stability of the trajectories in the presence of noise.
\end{abstract}

\section{Introduction}

The question under which conditions macroscopic matter takes crystalline form, having long-range periodic order at sufficiently low temperatures, is a long-standing open problem in statistical mechanics~\cite{Simon1984, Radin1987, BL2005, BL2015}. It was shown that generic interactions have ground states that are non-periodic, non-mixing but weakly mixing~\cite{Miekisz1988,MR1989,EM2020}.

Discovery of quasicrystals~\cite{SBG+1984} changed a direction of research. A major problem is now to explain the formation and thermodynamic stability of solids with long-range non-periodic order; see a book on mathematics of quasicrystals \cite{BG2013} and a recent short review \cite{Miekisz2025}. 

While the crystal problem is inherently about matter in a continuous space, and should arguably be addressed with continuum models (quantum or classical), the quasicrystal problem remains non-trivial and fascinating even if abstracted in discrete lattice-based classical models. It has been clear that the formation of quasicrystals is linked, in one way or another, to the existence of aperiodic tile sets~\cite{Radin1987, Radin1985, Radin1986, Miekisz1986, Miekisz1987, Miekisz1990, Miekisz1993,Miekisz2025}
--- finite sets of tiles with local matching constraints that can tile the entire plane but only in non-periodic ways~\cite{GS1987}.

Discovery of aperiodic tile sets goes back to Berger in 1966, who constructed such a tile set (over 20,000 tiles) and used it to prove the undecidability of the tiling problem~\cite{Berger1966}. Robinson later constructed a simpler example with~56 tiles~\cite{Robinson1971}.
Since then, numerous constructions using a variety of different tools and ideas have been found~\cite{GS1987, Mozes1989, Kari1996, DRS2012, JR2021}.

To understand the stability of quasicrystals at positive temperatures, it is natural to study tilings of the plane with an aperiodic tile set while allowing a small but positive density of errors. An aperiodic tile set that exhibits a strong form of stability in the presence of independent low-density errors has been constructed by Durand, Romashchenko, and Shen~\cite{DRS2012}.  Gayral and Sablik~\cite{GS2023} showed that a variant of Robinson's tile set has a weaker stability property against independent errors. Random fluctuations at low temperatures are however not independent. 

In the setting of lattice-gas models, the macroscopic equilibrium states are described by Gibbs measures (also called Gibbs states). In this setting, an abstract version of the quasicrystal problem can be informally formulated as follows:
\begin{quote}
    \emph{Does there exist a finite-range lattice-gas model whose extremal Gibbs states at low temperatures are random perturbations of non-periodic ground-state configurations in the support of some translation-invariant measure (the ground-state measure)?}
\end{quote}
Ideally, the random perturbations will exhibit a ``sea-island'' picture similar to the one in the two-dimensional Ising model at low temperature: disagreements with a ground-state configuration form sparse isolated islands in a single sea of agreements. Such Gibbs states will inevitably be non-periodic themselves. 

The above problem is still widely open.  Let us mention a few related results before presenting our construction.

Lattice-gas models with finite-range interactions based on non-periodic Robinson's tilings were introduced by Radin and Mi\c{e}kisz~\cite{Radin1985, Radin1986, Miekisz1986, Miekisz1987, Miekisz1990}.
It was shown that phase transitions may occur in such models. In particular, Mi\c{e}kisz~\cite{Miekisz1990} constructed a model in which for every~$n$, there is a temperature~$T_n$ such that at every temperature $T < T_n$, there exists a Gibbs state with a period bigger than~$n$.

In~\cite{EM1990}, van~Enter and Mi\c{e}kisz constructed a one-dimensional model with absolutely summable interactions and non-periodic Gibbs states that are random perturbations of the Thue-Morse sequence.
The same authors, together with Zahradník, constructed a three-dimensional model with exponentially decaying interactions and non-periodic Gibbs states~\cite{EMZ1998}.
The construction was based on stacking the Thue-Morse sequences along two extra dimensions.  It involved one-dimensional fast decaying interactions enforcing Thue-Morse sequences as one-dimensional ground-state configurations~\cite{GMR+1989}, and ferromagnetic Ising interactions forcing repetitions in the other two directions.

Recently, Chazottes, Kucherenko, and Quas provided a general construction of long-range models with the so-called freezing phase transitions on any prescribed subshift in any dimension~\cite{CKQ2025}.  In particular, given an arbitrary tile set (for instance, an aperiodic one), their construction yields a model whose equilibrium measures below a certain critical temperature are all supported on valid tilings. Note that Gibbs measures for models without hard constraints generally have full support. Hence, freezing phase transitions can occur only for interactions that are sufficiently long-ranged (in particular, not absolutely summable), for which the DLR theorem does not hold.  Long-range interactions are known to exhibit exotic ``non-physical'' behaviors~\cite{Israel1975,Jenkinson2006}.

The model presented in the current paper lives in four dimensions and has \emph{finite-range} interactions.  It admits ground-state configurations that are obtained by stacking valid tilings with an aperiodic set of tiles along two extra dimensions, and as such, resembles the stacking construction of van~Enter, Mi\c{e}kisz and Zahradn\'{i}k mentioned above.  The stability of these ground-state configurations at low temperature is, however, achieved differently, using results from the theory of fault-tolerant cellular automata (CA). The construction is based on three ingredients, which can be treated more or less as black boxes:
\begin{enumerate}[label={\arabic*)}]
    \item A method of simulating a CA with another CA with two extra dimensions that is resilient against noise, due to Toom, Gács and Reif~\cite{Toom1980, GR1988}.

    \item The existence of aperiodic sets of Wang tiles that are deterministic in one direction~\cite{Kari1992}.
        An example of such a tile set is Ammann's set of 16 tiles~\cite{GS1987, AGS1992}.

    \item The observation by Domany and Kinzel~\cite{DK1984} and developed by Goldstein et al.~\cite{GKL+1989}
        that distributions of the space-time trajectories of a positive-rate probabilistic CA (PCA) are Gibbs states for interactions corresponding to transition probabilities in PCA.
\end{enumerate}

The following section contains an informal description of the model.  The above three ingredients are described in Section~\ref{sec:ingredients}.  The construction is presented in Section~\ref{sec:construction}.  The main theorem is stated in Section~\ref{sec:main-theorem}.  More details on the first ingredient are provided in the Appendix A.

\section{Informal description}
We construct a 4-dimensional lattice-gas model with finite-range interactions between particles of 17 types, which occupy the sites of $\ZZ^{4}$. There are 16 types of particles corresponding to Ammann's tiles~\cite{GS1987, AGS1992}, which tile the plane only in a non-periodic way, and a ``blank'' particle.
There are uncountably many Ammann tilings, but they all look the same: every local patch of tiles occurs with the same frequency in all of them. This means that there exists a unique translation-invariant probability measure supported by them.

The interaction energies are all non-negative, and non-zero only for particle arrangements on translations of a $7$-element subset of the $2\times 2\times 2\times 2$ hypercube.

Every configuration that forms a valid tiling along the last two directions and is constant along the first two
is a configuration where all 7-particle interactions attain the same minimum and are therefore ground-state configurations. So are the all-blank configuration and configurations with interfaces between partial tilings and blank configurations. 

We refer to the ground-state configurations that are constant along the first two directions as \emph{standard}. There are uncountably many ground-state configurations that are not standard.
The standard ground-state configurations are stable at sufficiently low temperatures: there exist Gibbs states supported by configurations that are small perturbations of ground-state configurations satisfying the sea-island picture. We conjecture that there are no Gibbs states corresponding to non-standard ground-state configurations.

Let us note that it is easy to construct finite-range models with non-periodic Gibbs states. Consider, in three dimensions, uncoupled two-dimensional ferromagnetic Ising models stacked along the third dimension. Such a model trivially admits uncountably many extremal Gibbs states, corresponding to all one-dimensional sequences of pluses and minuses.
Such degenerate cases are arguably not of much interest, although a precise formulation of what should be considered as ``degenerate'' is beyond the scope of this article.  In the construction by van~Enter, Mi\c{e}kisz, and Zahradn\'{i}k~\cite{EMZ1998}, the Ising planes are coupled by exponentially decaying interactions that favour Thue-Morse sequences across planes.
Thanks to the infinite redundancy provided by the Ising planes, such exponentially decaying interactions are sufficient to ensure the survival of the Thue-Morse structure at sufficiently low temperatures.
The construction in the current paper, although it similarly employs stacking for redundancy, uses a different mechanism to achieve stability.

\section{Ingredients of the construction}
\label{sec:ingredients}

\subsection{Gács--Reif stacking construction}

G\'acs and Reif~\cite{GR1988} proposed a simple construction to simulate a CA with another one that is resilient against noise. The simulating CA requires two extra dimensions. It uses infinite redundancy to preserve information and Toom's rule~\cite{Toom1980} to correct errors.

To be specific, let $F\colon\Sigma^{\ZZ}\to\Sigma^{\ZZ}$ be a one-dimensional CA with a finite alphabet $\Sigma$. For concreteness (and since we will use such a form), we assume that $F$ has neighborhood $N=\{0,1\}$, so that
\begin{equation}
	(Fx)_i = f(x_i,x_{i+1})
\end{equation}
for some local rule $f\colon\Sigma^2\to\Sigma$.
The simulating CA $\tilde{F}\colon\Sigma^{\ZZ^3}\to\Sigma^{\ZZ^3}$ has the same alphabet $\Sigma$ but acts on three-dimensional configurations. For each $y\in\Sigma^{\ZZ^3}$, the image $\tilde{F}y$ is defined as
\begin{equation}
	(\tilde{F}y)_{a,b,i} \isdef f\big(y'_{(a,b,i)},y'_{(a,b,i+1)}\big)
\end{equation}
where $y'$ is given by Toom's North-East-Center (NEC) majority rule
\begin{equation}
\label{eq:gac-reif:correction}
	y'_{a,b,i} \isdef \majority\big(y_{(a,b,i)},y_{(a+1,b,i)},y_{(a,b+1,i)}\big) \;.
\end{equation}
In the case where $a,b,c\in\Sigma$ are distinct, $\majority(a,b,c)$ can be defined arbitrarily.
Thus, to apply~$\tilde{F}$, we first apply Toom's NEC rule on every plane $P_i\isdef\{(a,b,i): a,b,\in\ZZ\}$
and then apply the CA $F$ on every line $L_{a,b}\isdef\{(a,b,i): i\in\ZZ\}$.

Given a one-dimensional configuration $x\in\Sigma^{\ZZ}$,
we define its three-dimensional \emph{clone} $\kappa(x)\isdef\tilde{x}\in\Sigma^{\ZZ^3}$
by $\tilde{x}_{a,b,i}\isdef x_i$.
Observe that $\tilde{F}$ simulates $F$ in the sense that
for every $x\in\Sigma^{\ZZ}$, the diagram
\begin{align}
	{%
	\begin{tikzpicture}[yscale=0.6,>=stealth',baseline=(current bounding box.center)]
		\node (x) at (-1,1) {$x$};
		\node (y) at (1,1) {$Fx$};
		\node (xx) at (-1,-1) {$\tilde{x}$};
		\node (yy) at (1,-1) {$\tilde{F}\tilde{x}$};
		\draw[->] (x) -- node[above]{$F$} (y);
		\draw[->] (xx) -- node[above]{$\tilde{F}$} (yy);
		\draw[->] (x) -- node[right]{$\kappa$} (xx);
		\draw[->] (y) -- node[right]{$\kappa$} (yy);	
	\end{tikzpicture}
	}
\end{align}
commutes.  Note that the cloning map $x\mapsto\tilde{x}$ is one-to-one and it is trivial to recover a configuration~$x$
from its clone~$\tilde{x}$. The trajectory of~$\tilde{F}$ on a cloned configuration~$\tilde{x}$ thus has all the information about the trajectory of~$F$ on~$x$ with an infinite redundancy.

G\'acs and Reif showed that the trajectories of cloned configurations under $\tilde{F}$ are stable against noise in the same sense as Toom's CA, hence the simulation is reliable against noise.
To be precise, a random space-time configuration $X$ chosen from $\Sigma^{\ZZ^3\times\ZZ_{\geq 0}}$ is called an \emph{$\varepsilon$-perturbed} trajectory of $\tilde{F}$ if for every finite set $A\subseteq\ZZ^3\times\ZZ_{\geq 0}$, we have
\begin{equation}
    \PP\big(\text{$X$ violates the local rule at every $(k,t)\in A$}\big) \leq \varepsilon^{\abs{A}} \;.
\end{equation}
A special case is that of a \emph{uniformly} $\varepsilon$-perturbed trajectory, in which the symbol at each $(k,t)\in\ZZ^3\times\ZZ_{> 0}$ follows the local rule with probability~$1-\varepsilon$ and is chosen uniformly at random with probability~$\varepsilon$, independently of the other space-time positions.  In our construction, we only use such perturbations.
\begin{theorem}[G\'acs--Reif reliable simulation~\cite{GR1988}]
\label{thm:gacs-reif:stability}
	Let $F\colon\Sigma^{\ZZ}\to\Sigma^{\ZZ}$ be an arbitrary CA and $\tilde{F}$ the stacked version of $F$
	as described above.
	For every $\delta>0$, there exists an $\varepsilon>0$ such that for every configuration $x\in\Sigma^{\ZZ}$, if $X$ is any $\varepsilon$-perturbed trajectory of $\tilde{F}$ with a cloned initial configuration $\tilde{x}=\kappa(x)$, we have
	\begin{equation}
		\sup_{(\ell,s)\in\ZZ^3\times\ZZ_{\geq 0}} \PP\big(X_{\ell,s}\neq(\tilde{F}^s\tilde{x})_\ell\big) \leq \delta \;.
	\end{equation}
\end{theorem}


A similar simulation result, without the need for two extra dimensions but with a far more sophisticated encoding, was obtained by Gács~\cite[Theorem 5.8]{Gacs2001}.

Theorem~\ref{thm:gacs-reif:stability} follows from Toom's celebrated stability theorem~\cite{Toom1980}, which characterizes the monotone CA with binary alphabet for which, in the presence of noise, the all-$\symb{0}$ configuration remains stable. 
In addition to the remarkable construction, Gács and Reif gave a new proof of a quantitative version of Toom's theorem.  Other proofs of Toom's theorem (including simplifications and reformulations of the original one) have been offered~\cite{Toom1980,BS1988,GR1988,LMS1990,BG1991,MP2012,Ponselet2013,Gacs2021,SST2026}. 

Although not explicitly mentioned in the statement of the above theorem, its proof can also be used to show that, for sufficiently small~$\varepsilon$, the $\varepsilon$-perturbed trajectories satisfy the sea-island picture.  We say that a configuration $y$ has the \emph{range-$r$ sea-island picture} with respect to a configuration~$x$ if the set $\Delta(x,y)\isdef\{i:x_i\neq y_i\}$ of disagreements between $x$ and $y$ has no infinite range-$r$ clusters, and the complement of~$\Delta(x,y)$ has a unique infinite range-$1$ cluster. The \emph{range-$r$ clusters} of a set $A\subseteq\ZZ^{3+1}$ are the connected components of the graph with vertex set $A$ in which two vertices are adjacent if and only if they have lattice distance at most $r$.  For our purpose, we need the following version of the G\'acs--Reif result, which again follows from a corresponding statement in the setting of Toom's theorem:

\begin{manualtheorem}{\ref*{thm:gacs-reif:stability}$'$}[G\'acs--Reif reliable simulation: sea-island picture]
\label{thm:gacs-reif:stability:sea-island}
	Let $F\colon\Sigma^{\ZZ}\to\Sigma^{\ZZ}$ be an arbitrary CA and $\tilde{F}$ the stacked version of $F$
	as described above.
	For every $\delta>0$ and $r\in\NN$, there exists an $\varepsilon_0>0$ such that, for every $\varepsilon<\varepsilon_0$ and every bi-infinite cloned trajectory $\tilde{x}$ of $\tilde{F}$ (i.e., $\tilde{x}_{a,b,k,t}\isdef x_{k,t}$ where $x$ is a trajectory of~$F$), $\tilde{F}$ has a uniformly $\varepsilon$-perturbed trajectory $X$ such that
    \begin{enumerate}[label={\textup{(\alph*)}}]
        \item \label{claim:gacs-reif:enhanced:weak-stability}
            $\displaystyle \sup_{(\ell,s)\in\ZZ^3\times\ZZ}\PP\big(X_{\ell,s}\neq \tilde{x}_{\ell,s}\big) \leq \delta$.
        \item \label{claim:gacs-reif:enhanced:sea-island}
            $X$ has almost surely the range-$r$ sea-island picture with respect to~$\tilde{x}$.
    \end{enumerate}
\end{manualtheorem}

In Appendix~\ref{apx:toom-gacs-reif}, we describe the connection between the Gács--Reif result and Toom's stability theorem, and sketch a proof of Theorem~\ref{thm:gacs-reif:stability:sea-island} using one of the proofs of Toom's theorem.

\subsection{Cellular automaton based on deterministic Wang tiles}

A finite set of Wang tiles is said to be \emph{North-West deterministic} (NW-deterministic for short) if the colors on the top and left edges uniquely determine each tile. This notion was introduced by Kari, who constructed a NW-deterministic aperiodic tile set based on Robinson's construction and proved the undecidability of the tiling problem when restricted to NW-deterministic sets~\cite{Kari1992}. An early example due to Ammann~\cite{GS1987,AGS1992,Labbe2025} is in fact deterministic in two opposite directions and has only $16$ tiles (Fig.~\ref{fig:ammann:Wang}). Other aperiodic tile sets with varying degrees of determinism have been constructed~\cite{KP1999,Zinoviadis2016,Labbe2025}.  
In particular, the construction by Guillon and Zinoviadis~\cite{Zinoviadis2016} is deterministic in all but two opposite directions.

\begin{figure}[!ht]
	\begin{center}
		\begin{tikzpicture}[x=7pt,y=7pt,scale=2]
			\useasboundingbox (-5,-4) rectangle (5,4);
			
			\amtile{1}{(-3,3)}
			\amtile{2}{(-1,3)}
			\amtile{3}{(1,3)}
			\amtile{4}{(3,3)}
			
			\amtile{5}{(-3,1)}
			\amtile{6}{(-1,1)}
			\amtile{7}{(1,1)}
			\amtile{8}{(3,1)}
			
			\amtile{9}{(-3,-1)}
			\amtile{10}{(-1,-1)}
			\amtile{11}{(1,-1)}
			\amtile{12}{(3,-1)}
			
			\amtile{13}{(-3,-3)}
			\amtile{14}{(-1,-3)}
			\amtile{15}{(1,-3)}
			\amtile{16}{(3,-3)}
		\end{tikzpicture}
	\end{center}
\caption{%
    Ammann's aperiodic set of Wang tiles.
}\label{fig:ammann:Wang}
\end{figure}

Every deterministic tile set can be turned into a one-dimensional CA by introducing an extra tile and extending matching rules involving the enlarged set of tiles. To be specific, suppose that $\Sigma$ is a NW-deterministic tile set. Then, for every two tiles $a,b\in\Sigma$, there exists at most one tile
$\rho(a,b)\in\Sigma$ that is consistent with $a$ on its left and $b$ on its top.  Pick an extra symbol $\blank$, which we call \emph{blank}, and let $\overline{\Sigma}\isdef\Sigma\cup\{\blank\}$.
Now we may define a local rule
$\overline{\rho}\colon\overline{\Sigma}\times\overline{\Sigma}\to\overline{\Sigma}$, where
\begin{equation}
	\overline{\rho}(a,b) \isdef
		\begin{cases}
			\rho(a,b)	& \text{if $\rho(a,b)$ is defined,}\\
			\blank		& \text{otherwise.}
		\end{cases}
\end{equation}
The new rule $\overline{\rho}$ defines a one-dimensional CA $F\colon\overline{\Sigma}^{\ZZ}\to\overline{\Sigma}^{\ZZ}$ by
\begin{equation}
	(F x)_i \isdef \overline{\rho}(x_i, x_{i+1}) \;.
\end{equation}
This construction was used by Kari to prove the undecidability of the nilpotency problem for one-dimensional CA~\cite{Kari1992}.

Observe that every valid tiling by tiles in $\Sigma$ corresponds (up to an affine transformation of the plane) to a unique bi-infinite trajectory of the CA $F$ in which the blank symbol does not appear. In fact, this is a one-to-one correspondence: every bi-infinite trajectory with no blanks corresponds to a unique valid tiling. Another important observation is that the blank symbol spreads towards the left: if $x\in\overline{\Sigma}^{\ZZ}$ has a blank symbol at site $k$, then $Fx$ has blank symbols at sites $k-1$ and $k$,
and $F^n x$ has blanks all over the interval $[k-n,k]$.
In particular, if $\tilde{x}$ is a bi-infinite space-time trajectory of $F$
and $\tilde{x}_{k,t}$ is blank, then the entire cone
$C_{k,t}\isdef\{(\ell,s): \text{$s\geq t$ and $k-t+s\leq \ell\leq k$}\}$
is blank (Fig.~\ref{fig:ammann:CA}).
Clearly, the all-blank configuration $\unif{\blank}\in\overline{\Sigma}^{\ZZ^d}$
is a bi-infinite trajectory of the CA.

\begin{figure}[!ht]
	\begin{center}
		{%
		\begin{tikzpicture}[x=10pt,y=10pt,scale=2,>=stealth',rotate=-45]
			
			\amtile{1}{(-4,-4)}\amtile{2}{(-3,-3)}\amtile{5}{(-2,-2)}\amtile{13}{(-1,-1)}
			\amtile{12}{(0,0)}
			\amtile{8}{(1,1)}\amtile{4}{(2,2)}\amtile{10}{(3,3)}\amtile{16}{(4,4)}
			
			\amtile{10}{(-3,-4)}\amtile{3}{(-2,-3)}\amtile{16}{(-1,-2)}\amtile{4}{(0,-1)}
			\amtile{12}{(1,0)}
			\amtile{7}{(2,1)}\amtile{6}{(3,2)}\amtile{1}{(4,3)}
			
			\amtile{4}{(-3,-5)}\amtile{11}{(-2,-4)}\amtile{16}{(-1,-3)}\amtile{7}{(0,-2)}
			\amtile{6}{(1,-1)}
			\amtile{9}{(2,0)}\amtile{3}{(3,1)}\amtile{15}{(4,2)}\amtile{12}{(5,3)}
			
			\amtile{5}{(-2,-5)}\amtile{1}{(-1,-4)}\xblank{(0,-3)}\amtile{3}{(1,-2)}
			\amtile{2}{(2,-1)}
			\amtile{11}{(3,0)}\amtile{9}{(4,1)}\amtile{4}{(5,2)}
			
			\amtile{3}{(-2,-6)}\amtile{14}{(-1,-5)}\xblank{(0,-4)}\xblank{(1,-3)}
			\xblank{(2,-2)}
			\amtile{3}{(3,-1)}\xblank{(4,0)}\xblank{(5,1)}\amtile{15}{(6,2)}
			
			\amtile{16}{(-1,-6)}\xblank{(0,-5)}\xblank{(1,-4)}\xblank{(2,-3)}
			\xblank{(3,-2)}
			\xblank{(4,-1)}\xblank{(5,0)}\xblank{(6,1)}
			
			\amtile{1}{(-1,-7)}\xblank{(0,-6)}\xblank{(1,-5)}\xblank{(2,-4)}
			\xblank{(3,-3)}
			\xblank{(4,-2)}\xblank{(5,-1)}\xblank{(6,0)}\xblank{(7,1)}
			
			\xblank{(0,-7)}\xblank{(1,-6)}\xblank{(2,-5)}\xblank{(3,-4)}
			\xblank{(4,-3)}
			\xblank{(5,-2)}\xblank{(6,-1)}\xblank{(7,0)}
			
			\xblank{(0,-8)}\xblank{(1,-7)}\xblank{(2,-6)}\xblank{(3,-5)}
			\xblank{(4,-4)}
			\xblank{(5,-3)}\xblank{(6,-2)}\xblank{(7,-1)}\xblank{(8,0)}
			
			\begin{scope}[overlay]
				\draw[thick,->] (-4.8,-7) to node[left]{\scriptsize time} (-3.2,-8.6);
			\end{scope}
			\draw[thick,->] (-1,-1)+(-1.5,0) -- ++(-0.6,0);
			\draw[thick,->] (0,0)+(-1.5,0) -- ++(-0.6,0);
			\draw[thick,->] (1,1)+(-1.5,0) -- ++(-0.6,0);
			\node[fill=white,inner sep=1pt] at (-2.7,-1) {\scriptsize $k\!-\!1$};
			\node[fill=white,inner sep=1pt] at (-1.7,0) {\scriptsize $k$};
			\node[fill=white,inner sep=1pt] at (-0.7,1) {\scriptsize $k\!+\!1$};
		\end{tikzpicture}
		}
	\end{center}
\caption{%
	A sample space-time trajectory of the CA associated with Ammann's aperiodic Wang tiles.
	The diagram is tilted to make it look like a tiling.
    Circles indicate blanks.
}\label{fig:ammann:CA}
\end{figure}

\subsection{PCA and space-time Gibbs states}

Domany and Kinzel~\cite{DK1984} observed that the random space-time trajectories of a $d$-dimensional PCA with positive transition probabilities
(i.e., at every local update, every symbol has a non-zero probability of appearing) are distributed according to Gibbs states for an associated finite-range interaction.
Goldstein et al.~\cite{GKL+1989} showed that the converse is also true if we restrict ourselves to translation-invariant measures: every translation-invariant Gibbs state for the associated interaction is the distribution of a stationary random space-time trajectory of the PCA. 

\begin{theorem}[PCA trajectories vs.\ Gibbs states~\cite{DK1984,GKL+1989}]
\label{thm:PCA-vs-Gibbs}
    Let $\Psi$ be a $d$-dimensional PCA with finite neighborhood~$N\Subset\ZZ^d$ and a strictly positive transition rule~$\psi:\Sigma^N\times\Sigma\to[0,1]$.
    The distribution of every bi-infinite random trajectory of $\Psi$ is a $(d+1)$-dimensional Gibbs state \textup{(}at the inverse temperature $1$\textup{)}
	for the interactions given by the observable $\varphi(x)\isdef-\log\psi(x_{N\times\{-1\}},x_{0,0})$.
	Conversely, every translation-invariant Gibbs state for~$\varphi$ is the distribution of a bi-infinite random trajectory of~$\Psi$.
\end{theorem}
Here, the observable $\varphi$ describes the energy contribution of the space-time position~$(\uline{0},0)$, which depends only on the pattern seen on~$\tilde{N}\isdef\{(\uline{0},0)\}\cup\{(a,-1): a\in N\}\subseteq\ZZ^d\times\ZZ$.  The energy contribution of the space-time position $(k,t)\in\ZZ^d\times\ZZ$ is the application of $\varphi$ on the space-time configuration shifted in space by~$k$ and in time by~$t$, thus depends only on the pattern seen on~$(k,t)+\tilde{N}$.

In general, the above correspondence may be lost when the temperature is varied.  However, as we shall see in the next section, for the PCA that are perturbations of deterministic CA with independent uniform noise, there is a simple correspondence between temperature and the noise parameter.

\section{Construction: combining the above ingredients}
\label{sec:construction}

We now construct a 4-dimensional finite-range lattice-gas model that has non-degenerate non-periodic ground-state configurations that are stable at positive temperatures.

\begin{enumerate}[label={(\roman*)}]
	\item We start with Ammann's aperiodic set of Wang tiles.
		Let $\Sigma$ denote the set of tiles.
	\item We turn this tile set into a CA $F$ by adding an extra blank symbol $\blank$.
		The CA $F$ has the symbol set $\overline{\Sigma}\isdef\Sigma\cup\{\blank\}$
		and the local rule $f:\overline{\Sigma}\times\overline{\Sigma}\to\overline{\Sigma}$.
	\item We simulate $F$ by a three-dimensional noise-resilient CA $\tilde{F}$
		given by the stacking construction of G\'acs and Reif.
		The CA $\tilde{F}$ has neighborhood
		\begin{equation}
			N \isdef \{(0,0,0), (1,0,0), (0,1,0), (0,0,1), (1,0,1), (0,1,1)\} \;,
		\end{equation}
		and its local rule is denoted by $\tilde{f}:\overline{\Sigma}^N\to\overline{\Sigma}$.
	\item We add a uniform noise to $\tilde{F}$ to build
		a PCA $\Psi_\varepsilon$. 
		Namely, for $0<\varepsilon<1$,
		we let $\theta_\varepsilon:\overline{\Sigma}\times\overline{\Sigma}\to[0,1]$
		be the stochastic matrix defined by
		\begin{align}
			\theta_\varepsilon(a,b) &\isdef
				\begin{dcases}
					1-\varepsilon	& \text{if $b=a$,}\\
					\varepsilon/(\abs{\smash{\overline{\Sigma}}}-1)	& \text{otherwise}
				\end{dcases}
		\end{align}
		We define the local rule $\psi_\varepsilon:\overline{\Sigma}^N\times\overline{\Sigma}\to[0,1]$
		as $\psi_\varepsilon(p,b)\isdef\theta_\varepsilon(\tilde{f}(p),b)$.
		Observe that this PCA has strictly positive transition probabilities.  Furthermore, every random trajectory of $\Psi_\varepsilon$ is a uniformly $\varepsilon$-perturbed trajectory of~$\tilde{F}$, so Theorems~\ref{thm:gacs-reif:stability} and~\ref{thm:gacs-reif:stability:sea-island} apply.
	\item We define a 4-dimensional lattice-gas model with
		configurations in $\overline{\Sigma}^{\ZZ^4}$
		and interaction
		\begin{align}
			\varphi_\varepsilon(y) &\isdef -\log\psi_\varepsilon(y_{N\times\{-1\}},y_{0,0})
		\end{align}
		as suggested by Theorem~\ref{thm:PCA-vs-Gibbs}.
		Note that $\varphi_\varepsilon(y)$ depends only on the restriction of $y$ to the finite (seven-element) set
        \begin{equation}
            \tilde{N} \isdef \{(0,0,0,-1), (1,0,0,-1), (0,1,0,-1), (0,0,1,-1), (1,0,1,-1), (0,1,1,-1), (0,0,0,0)\} \;.
        \end{equation}
\end{enumerate}

Observe that, for every space-time configuration $y$, the value of $\varphi_\varepsilon(y)$ depends only on whether $y$ has an error at the origin or not (i.e., whether $y$ follows the local rule $\tilde{f}$ at the space-time position $(0,0)$ or not).  More specifically,
\begin{equation}
    \varphi_\varepsilon(y) \isdef
        \begin{cases}
            -\log(1-\varepsilon)        & \text{if $y$ is error-free at the origin,} \\
            -\log(\varepsilon/16) & \text{if $y$ has an error at the origin.}
        \end{cases} 
\end{equation}
The promised lattice-gas model is the one with interactions given by
\begin{equation}
    \varphi(y) \isdef
        \begin{cases}
            0           & \text{if $y$ is error-free at the origin,} \\
            1           & \text{if $y$ has an error at the origin.}
        \end{cases} 
\end{equation}
Note that, for $\varepsilon_0\isdef 16/(16+\ee)$, we have $-\log(\varepsilon_0/16)=-\log(1-\varepsilon_0)+1$, thus $\varphi$ is equivalent to~$\varphi_{\varepsilon_0}$, in the sense that $\varphi$ and $\varphi_{\varepsilon_0}$ have the same Gibbs measures at any temperature.

At inverse temperature~$\beta>0$, the model with interaction~$\varphi_{\varepsilon_0}$ is equivalent to the model with interaction $\beta\varphi_{\varepsilon_0}$ at inverse temperature~$1$.  
The interaction $\beta\varphi_{\varepsilon_0}$ is, in turn, equivalent to the interaction~$\varphi_\varepsilon$ associated with the PCA rule $\psi_\varepsilon$ with a noise parameter $\varepsilon=\varepsilon(\beta)$ depending on~$\beta$.
Namely, solving $\beta\varphi_{\varepsilon_0}=\alpha+\varphi_\varepsilon$ for $\alpha$ and $\varepsilon$ gives the unique solution
\begin{equation}
    \alpha\isdef-\log\big[(1-\varepsilon_0)^\beta+16(\varepsilon_0/16)^\beta\big] \;,
    \qquad\qquad
    \varepsilon\isdef \frac{16(\varepsilon_0/16)^\beta}{(1-\varepsilon_0)^\beta+16(\varepsilon_0/16)^\beta} \;.
\end{equation}
Since $\varepsilon_0<16/17$, $\varepsilon(\beta)$ is strictly decreasing.
Furthermore, $\varepsilon(\beta)\to 0$ as $\beta\to\infty$, which means low temperature corresponds to low noise.


Observe that, for every bi-infinite trajectory~$z$ of the CA~$F$, the lattice-gas model with interactions given by~$\varphi$ ($\equiv\varphi_{\varepsilon_0}$) has a ground-state configuration that is the clone of~$z$ (i.e., obtained by stacking~$z$ along two extra dimensions).  Combining Theorems~\ref{thm:gacs-reif:stability:sea-island} and Theorem~\ref{thm:PCA-vs-Gibbs}, we find that any such ground-state configuration is stable at positive temperatures.

More specifically, let $z$ be a bi-infinite trajectory of $F$
where $z_{k,t}$ denotes the symbol at site $k$ and time $t$. Let $\tilde{z}$ denote the cloned trajectory of $\tilde{F}$
defined by $\tilde{z}_{a,b,k,t}\isdef z_{k,t}$.
It follows from Theorem~\ref{thm:gacs-reif:stability:sea-island} that, for each $\delta>0$ and $r\in\NN$, if we choose $\varepsilon>0$ sufficiently small, then the PCA $\Psi_\varepsilon$ has a bi-infinite random trajectory~$Z$
that almost surely has the range-$r$ sea-island picture with respect to~$\tilde{z}$ and satisfies
\begin{equation}
	\sup_{(\ell,s)\in\ZZ^3\times\ZZ} \PP\big(Z_{\ell,s}\neq \tilde{z}_{\ell,s}\big) < \delta \;.
\end{equation}
It follows from Theorem~\ref{thm:PCA-vs-Gibbs} that when $\beta$ is sufficiently large, that is $\varepsilon$ is sufficiently small, the lattice-gas model with interaction~$\varphi$ has a Gibbs state $\mu_\beta$ at inverse temperature $\beta$ that assigns probability~$1$ to the configurations that have the range-$r$ sea-island picture with respect to~$\tilde{z}$ and satisfies
\begin{equation}
	\inf_{a\in\ZZ^4} \mu_\beta(\{x \in \Omega, x_a = \tilde{z}_a\}) \geq 1-\delta \;.
    \label{eq:high-prob-agreement}
\end{equation}

The cloned bi-infinite trajectories of~$F$ are precisely the standard ground-state configurations of the model.
They include the cloned Ammann tilings, as well as the all-blank configuration and configurations that agree with a cloned Ammann tiling on one side and are blank on the other.
All such ground-state configurations are stable in the above sense.

However, we cannot rule out the stability of other non-standard ground configurations, or the possibility of other Gibbs measures at low temperature that are not perturbations of bi-infinite trajectories.  A complete characterization of bi-infinite random trajectories of the PCA $\Psi_{\varepsilon}$ is missing. 

\section{Quasicrystalline Gibbs states at low temperatures}
\label{sec:main-theorem}

Let $z$ be a standard ground-state configuration corresponding to an Ammann tiling, and let $\mu_\beta$ be a Gibbs state at a sufficiently small temperature $T=1/\beta$ that is $\delta$-close to $z$. Let us show that $\mu_\beta$ is non-periodic.

\begin{theorem}[Non-periodic Gibbs measures]
Suppose that $\delta<\nicefrac{1}{2}$.
Then, the Gibbs state $\mu_\beta$ is non-periodic along the last two directions.
\end{theorem}

\begin{proof}
The assumption $\delta<\nicefrac{1}{2}$ ensures that for every $k\in\ZZ^4$, there is at most one symbol $s\in\overline{\Sigma}$ such that $\mu_\beta(\{x: x_k=s\})\geq 1-\delta$.

Let us assume that $p=(p_1,p_2,p_3,p_4)\in\ZZ^4$ is a period of $\mu_\beta$.
Then, for every $k\in\ZZ^4$, we have
\begin{align}
    \mu_\beta(\{x: x_{k+p}=z_k\})=\mu_\beta(\{x:x_k=z_k\}) &\geq 1-\delta \;, \\
    \mu_\beta(\{x:x_{k+p}=z_{k+p}\}) &\geq 1-\delta \;,
\end{align}
for which it follows $z_{k+p}=z_k$.
Therefore, $z$ has period~$p$.
This contradicts the non-periodicity of the Ammann tilings unless $p_3=p_4=0$.
%
%
\end{proof}

We can now formulate our main theorem, which follows from the discussion in the previous section.

\begin{theorem}[Quasicrystalline Gibbs states]
For every $\delta>0$ and $r\in\NN$, there is a $\beta_0>0$ such that, for every $\beta\geq\beta_0$ and every standard ground-state configuration $z$ corresponding to a non-periodic Ammann tiling, there exists a Gibbs state $\nu_z^{(\beta)}$ such that
\begin{enumerate}[label={\textup{(\alph*)}}]
		\item $\nu_z^{(\beta)}$ is uniformly $\varepsilon$-close to $z$, that is,
			\begin{equation}
				\sup_{k\in Z^4} \nu_z^{(\beta)}\big(x: x_k\neq z_k\big) < \delta \;.
			\end{equation}
		\item Under $\nu_z^{(\beta)}$, almost every configuration $x$ has the range-$r$ sea-island picture
			with respect to $z$, that is, 
            the disagreements of $x$ and $z$ do not have any infinite, range-$r$ clusters.
	\end{enumerate}
\end{theorem}

\section{Discussion} 

We have constructed a 4-dimensional lattice-gas model with finite-range interactions and with ground-state configurations which correspond to non-periodic Ammann tilings (they are constant in two other directions). We showed that, at low temperatures, there exist non-periodic Gibbs measures that are small perturbations of such ground-state configurations. However, our lattice-gas model also has many other ground-state configurations, including a translation-invariant homogeneous ground-state configuration consisting of just blank particles and the corresponding translation-invariant extremal Gibbs measure.  We conjecture that we can eliminate such a Gibbs measure by introducing an arbitrarily small chemical potential assigning a positive energy to a blank particle.
Introducing such a chemical potential would, however, break the correspondence with PCA, hence the proof may require other techniques.
We also conjecture that the non-standard ground-state configurations (i.e., the ones that are not constant in the two extra directions) do not give rise to Gibbs measures.

The existence of a 3-dimensional lattice-gas model with finite-range interactions, a unique ground-state measure supported by non-periodic ground-state configurations, and with non-periodic extremal Gibbs measures that are small perturbations of non-periodic ground-state configurations remains a fundamental problem in statistical physics.

We remark that the one-dimensional simulation result by Gács gives rise to a two-dimensional lattice-gas model with finite-range interactions that has uncountably many non-periodic extremal Gibbs states~\cite{Gacs2001}. Gács's construction is, however, very sophisticated and requires an astronomical number of symbols.
The construction presented in the current paper has the advantage of being simple and concrete.

\paragraph{Acknowledgments.}
We thank Aernout van~Enter for numerous insightful discussions.
{S.T.} also thanks Peter Gács for helpful discussions.
We thank the National Science Centre (Poland) for financial support under Grant No.~016/22/M/ST1/00536 and the IDUB Thematic Research Programme ``Excellence Initiative - Research University (2020-2025)'' at the University of Warsaw -``Quasicrystals mathematical physics, ergodic theory, and topology of nonperiodic structures'' during which final results were obtained.  S.~Taati gratefully acknowledges the support of the Center for Advanced Mathematical Sciences (CAMS) at the American University of Beirut (CAMS ORCID: 0009-0004-5763-5004).

\paragraph{Data Availability.} This manuscript has no associated data.

\paragraph{Conflict of interest.} The authors declare that there is no conflict of interest.

\bibliographystyle{plainurl}
\bibliography{bibliography}

\appendix

\section*{Appendix}

\section{Sea-island picture in Gács--Reif simulation}
\label{apx:toom-gacs-reif}
The purpose of this appendix is to sketch the proof of Theorem~\ref{thm:gacs-reif:stability:sea-island} by first reducing it to a similar statement regarding one of the Toom models, and then adapting one of the available proofs of Toom's theorem.

\subsection{Gács--Reif simulation via Toom stability}
\label{apx:gacs-reif-via-toom}
Recall the notation $\Delta(x,y)\isdef\{i: x_i\neq y_i\}$ for the disagreement set of two configuraitons $x$ and $y$.  With some abuse of notation, we identify $\Delta(x,y)$ with a $\{\symb{0},\symb{1}\}$-configuration with $\Delta_i(x,y)\isdef\symb{1}$ if and only if $x_i\neq y_i$.  We denote by $\preceq$ the usual product partial order on configurations induced by $\symb{0}\prec\symb{1}$.
We denote the all-$\symb{0}$ configuration (in any dimension) by~$\unifO$.

Let us introduce a CA with binary alphabet that ``tracks'' the spread of disagreements under the Gács--Reif~CA.
Define a CA $D\colon \{\symb{0},\symb{1}\}^{\ZZ^3}\to\{\symb{0},\symb{1}\}^{\ZZ^3}$ as follows.  For each $z\in \{\symb{0},\symb{1}\}^{\ZZ^3}$, the image $Dz$ is defined by
\begin{equation}
\label{eq:our-toom-model}
    (Dz)_{a,b,i} \isdef
        \majority\big(z_{(a,b,i)}, z_{(a+1,b,i)}, z_{a,b+1,i}\big) \lor
        \majority\big(z_{(a,b,i+1)}, z_{(a+1,b,i+1)}, z_{a,b+1,i+1}\big) \;,
\end{equation}
where $\lor$ stands for logical disjunction.  This CA is in fact the Gács--Reif simulation of Stavskaya's~CA.
\begin{observation}[Tracking disagreements in deterministic trajectories]
    Let $x\in\Sigma^\ZZ$ be an arbitrary configuration and $\tilde{x}\isdef\kappa(x)$ its three-dimensional clone.  Let $y\in\Sigma^{\ZZ^3}$ be any other three-dimensional configuration.  Then 
    \begin{equation}
        \Delta(\tilde{F}\tilde{x}, \tilde{F}y)\preceq D\Delta(\tilde{x},y) \;.
    \end{equation}
\end{observation}
\noindent Since $\tilde{F}\tilde{x}$ is the clone of $Fx$, it follows by induction that
$\Delta(\tilde{F}^t\tilde{x}, \tilde{F}^ty)\preceq D^t\Delta(\tilde{x},y)$
for every $t\geq 0$.

The above relationship between $D$ and $\tilde{F}$ can be extended to random perturbations.
Since the symbol $\symb{1}$ represents a disagreement, we focus our attention to one-sided perturbed trajectories of~$D$.
A \emph{one-sided} $\varepsilon$-perturbed trajectory of~$D$ is an $\varepsilon$-perturbed trajectory $X$ that always follows the rule when the rule dictates a~$\symb{1}$.  (In other words, errors can occur only when the local rule dictates a~$\symb{0}$.)
A \emph{uniform} one-sided $\varepsilon$-perturbed trajectory of $D$ is a random trajectory of the PCA $D_\varepsilon$ obtained by adding one-sided noise $\symb{0}\overset{\varepsilon}{\to}\symb{1}$ to~$D$.
\begin{lemma}[Tracking disagreements in random perturbations]
\label{lem:tracking-disagreements:perturbations}
    Let $x\in\Sigma^\ZZ$ be an arbitrary configuration and $\tilde{x}\isdef\kappa(x)$ its three-dimensional clone.
    Let $Y$ be an $\varepsilon$-perturbed trajectory of $\tilde{F}$ with initial configuration~$\tilde{x}$.
    Then, $Y$ can be coupled with a one-sided $\varepsilon$-perturbed trajectory $Z$ of~$D$ with initial configuration~$\unifO$ such that $\Delta\big((\tilde{F}^t\tilde{x})_{t\geq 0}, Y\big)\preceq Z$ almost surely.
    Furthermore, if $Y$ is a uniform perturbation, $Z$ can also be chosen to be a uniform one-sided perturbation.
\end{lemma}
\begin{proof}
    Let us denote the local rule of $D$ by~$\rho$.  Note that $\rho$ has the same neighborhood as~$\tilde{f}$, which as before, we denote by~$\tilde{N}$.

    Let $Y$ be an $\varepsilon$-perturbed trajectory of~$\tilde{F}$ starting with~$\tilde{x}$.
    Let $E$ be the random space-time configuration in~$\{\symb{0},\symb{1}\}^{\ZZ^3\times\ZZ_{\geq 0}}$ indicating the positions at which~$Y$ has an error, that is,
    \begin{equation}
        E_{k,t} \isdef
            \begin{cases}
                \symb{1}    & \text{if $t>0$ and $Y_{k,t}\neq \tilde{f}(Y_{k+\tilde{N},t-1})$,} \\
                \symb{0}    & \text{otherwise.}
            \end{cases}
    \end{equation}
    Construct $Z$ recursively by setting $Z_{k,0}\isdef\symb{0}$ for every $k\in\ZZ^3$ and
    \begin{equation}
        Z_{k,t} \isdef \rho\big(Z_{k+\tilde{N},t-1}\big)\lor E_{k,t}
    \end{equation}
    for $k\in\ZZ^3$ and $t>0$.
    Clearly, $Z$ is a one-sided $\varepsilon$-perturbed trajectory of~$D$ starting with~$\unifO$, and that $\Delta\big((\tilde{F}^t\tilde{x})_{t\geq 0}, Y\big)\preceq Z$.

    To see that the second claim holds, note that if $Y$ is a uniform perturbation, the configuration $E$ is~i.i.d., hence $Z$ is also a uniform one-sided perturbation.  (To be precise, $Z$ defined as such will have parameter $\PP(E_{k,t}=\symb{1})=\varepsilon(1-1/\abs{\Sigma})$ instead of~$\varepsilon$.  To remedy this, we can couple $E$ with another i.i.d.~configuration~$E'$ with $\PP(E'_{k,t}=\symb{1})=\varepsilon$ such that $E\preceq E'$ almost surely, and use $E'$ instead of~$E$ to construct~$Z$.)
\end{proof}

Let us now observe two key properties of~$D$.
\begin{itemize}
    \item First, $D$ is \emph{monotone} in the sense that $D(x)\preceq D(y)$ if $x\preceq y$.
    \item Second, $D$ has the \emph{erosion} property on~$\unifO$ (equivalent terminology: $D$ is a \emph{$\unifO$-eroder}).  This means that, for every configuration~$z$ that has only finitely many~$\symb{1}$s, there exists a time~$t\geq 0$ such that $D^t(x)=\unifO$.  Indeed, consider the prism
    \begin{equation}
        R_{m,\ell} \isdef \{(i,j,k)\in\ZZ^3: \text{$i\geq 0$ and $j\geq 0$ and $i+j\leq m$ and $-\ell\leq k\leq 0$}\} \;.
    \end{equation}
    It is easy to verify that if $\Delta(\unifO,x)\subseteq R_{m,\ell}$, then $\Delta(\unifO,Dx)\subseteq R_{m-1,\ell+1}$.  
    In particular, $\Delta(\unifO,D^{m+1} x)=\varnothing$, which means $D^{m+1} x=\unifO$.
    Now, if $x$ is an arbitrary configuration with finitely many~$\symb{1}$s, we can find $(a,b,c)\in\ZZ^3$ and $m,\ell\in\ZZ$ such that $\Delta(\unifO,x)\subseteq (a,b,c)+R_{m,\ell}$, and the claim follows from the translation symmetry of~$D$.
\end{itemize}

Theorem~\ref{thm:gacs-reif:stability} now becomes an immediate consequence of Toom's theorem:
\begin{theorem}[Toom's stability theorem~\cite{Toom1980}]
\label{thm:toom:stability}
	Let $d\geq 1$, and let $D\colon\{\symb{0},\symb{1}\}^{\ZZ^d}\to\{\symb{0},\symb{1}\}^{\ZZ^d}$ be a monotone CA that has the erosion property on~$\unifO$.  Then, $\unifO$ is weakly stable against random perturbations of~$D$ in the sense that, for every $\delta>0$, there exists an $\varepsilon>0$ such that if $Z$ is any one-sided $\varepsilon$-perturbed trajectory of~$D$ with initial configuration~$\unifO$, we have
	\begin{equation}
		\sup_{(\ell,s)\in\ZZ^3\times\ZZ_{\geq 0}} \PP\big(Z_{\ell,s}=\symb{1}\big) \leq \delta \;.
	\end{equation}
\end{theorem}
For monotone CA, the above stability property is in fact characterized by the erosion property~\cite{Toom1980}.  We do not need this fact.

\subsection{Sea-island picture in the Toom models}
We now use the above connection between the Gács--Reif CA $\tilde{F}$ and the associated monotone eroder~$D$ to sketch a proof of Theorem~\ref{thm:gacs-reif:stability:sea-island}.

Let $\delta>0$.  We set $\varepsilon_0\isdef\min\{\varepsilon_1,\varepsilon_2\}$, where $\varepsilon_1$ and $\varepsilon_2$ will be chosen below.

Choose $\varepsilon_1>0$ such that every $\varepsilon_1$-perturbed trajectory of~$\tilde{F}$ is $\delta$-close to an unperturbed trajectory of~$\tilde{F}$ in the sense of Theorem~\ref{thm:gacs-reif:stability}.
Note that for $\varepsilon<\varepsilon_1$, every $\varepsilon$-perturbed trajectory is also an $\varepsilon_1$-perturbed trajectory.

Let $x$ be a bi-infinite trajectory of~$F$ and $\tilde{x}$ its corresponding cloned trajectory of~$\tilde{F}$ with $\tilde{x}_{a,b,k,t}=x_{k,t}$ as in Theorem~\ref{thm:gacs-reif:stability:sea-island}.
The claimed bi-infinite perturbed trajectory~$X$ corresponding to~$\tilde{x}$ is constructed by taking a subsequential limit in a standard manner.  Namely, for $s\in\ZZ$, let $X^{(s)}$ be a random space-time configuration that almost surely agrees with $\tilde{x}$ at any time $t\leq s$ and forms a uniform $\varepsilon$-perturbed trajectory of~$\tilde{F}$ for $t\geq s$.  By compactness, there is a sequence $0>s_1>s_2>\cdots$ for which the distribution of $X^{(s_i)}$ converges weakly to a measure~$\mu$.  It is then clear that $\mu$ is the distribution of a bi-infinite uniform $\varepsilon$-perturbed trajectory~$X$ of~$\tilde{F}$, and that
\begin{equation}
    \sup_{(\ell,s)\in\ZZ^3\times\ZZ}\PP\big(X_{\ell,s}\neq \tilde{x}_{\ell,s}\big) \leq \delta \;.
\end{equation}
This establishes claim~\ref{claim:gacs-reif:enhanced:weak-stability} of the theorem.

Using Lemma~\ref{lem:tracking-disagreements:perturbations} and a similar limiting procedure as above, we can find a bi-infinite perturbed trajectory $Z$ of $D$ such that $\Delta(\tilde{x},X)\preceq Z$ almost surely and
\begin{equation}
    \sup_{(\ell,s)\in\ZZ^3\times\ZZ} \PP\big(Z_{\ell,s}=\symb{1}\big) \leq \delta \;.
\end{equation}
More specifically, using Lemma~\ref{lem:tracking-disagreements:perturbations}, we can couple $X^{(s)}$ with a random space-time $\{\symb{0},\symb{1}\}$-configuration $Z^{(s)}$ such that
\begin{itemize}
    \item $Z^{(s)}_{a,b,k,t}=\symb{0}$ for~$t\leq s$ and $Z^{(s)}$ forms a uniform one-sided~$\varepsilon$-perturbed trajectory of~$D$ for $t\geq s$.
    \item $\Delta(\tilde{x},X^{(s)})\preceq Z^{(s)}$.
\end{itemize}
A standard argument using monotonicity can now be used to show that, in this case, the distribution of $Z^{(s)}$ converges as $s\to-\infty$, even without taking a subsequence, to the distribution of a bi-infinite uniform one-sided $\varepsilon$-perturbed trajectory~$Z$ with the above property.  Furthermore, $Z$ is the smallest such trajectory in the sense that, any other bi-infinite uniform one-sided $\varepsilon$-perturbed trajectory~$Z'$ of~$D$ can be coupled with $Z$ in such a way that $Z\preceq Z'$ almost surely.

Claim~\ref{claim:gacs-reif:enhanced:sea-island} of the theorem now becomes a consequence of the following:
\begin{theorem}[Sea-island picture in Toom models]
\label{thm:toom:sea-island}
	Let $d\geq 1$, and let $D\colon\{\symb{0},\symb{1}\}^{\ZZ^d}\to\{\symb{0},\symb{1}\}^{\ZZ^d}$ be a monotone CA that has the erosion property on~$\unifO$.
    For every $r\in\NN$, there exists an $\varepsilon_2>0$, such that, for every $\varepsilon<\varepsilon_2$, the smallest (in the above sense) bi-infinite uniform one-sided $\varepsilon$-perturbed trajectory~$Z$ of~$D$ has almost surely the range-$r$ sea-island picture with respect to~$\unifO$.
\end{theorem}




Let us now sketch a proof of the latter theorem, following the idea of Gács and Reif~\cite{GR1988}.  The proof by Bramson and Gray~\cite{BG1991} is based on a similar idea, though it is implemented differently.

The proof of Theorem~\ref{thm:toom:stability} by Gács and Reif is based on the following ideas:
\begin{enumerate}[label={(\arabic*)}]
    \item In a monotone eroder, the erosion of a finite region of~$\symb{1}$s in a background of~$\symb{0}$s can be tracked by polyhedra that shrink with constant speed, similarly to the prisms~$R_{m,\ell}$ introduced in the previous subsection.
    \item In a noisy background, in space-time regions where the noise is sufficiently ``sparse'', a similar erosion still occurs and can be tracked by polyhedra that shrink at a slightly lower speed.  Here, the level of sparseness should be sufficient in comparison to the size of the polyhedra.
    \item In a uniform $\varepsilon$-perturbation with small $\varepsilon>0$, the set $E\subseteq\ZZ^d\times\ZZ$ of space-time positions $(k,t)$ at which an error occurs (i.e., a~$\symb{0}$ is turned into a~$\symb{1}$ due to noise) is almost surely ``sparse'' in a hierarchical sense.  Occasional bursts of noise that violate sparseness at a certain scale are themselves sparse at a higher scale.
\end{enumerate}
Theorem~\ref{thm:toom:sea-island} then follows after noting that
\begin{enumerate}[label={(\arabic*)}, resume]
    \item The space-time positions at which the smallest one-sided $\varepsilon$-perturbed trajectory of~$D$ has $\symb{1}$s is still almost surely ``sparse'', albeit in a slightly weaker sense.
    \item Every ``sparse'' subset of $\ZZ^d\times\ZZ$ has the sea-island picture.
\end{enumerate}

Using shrinking polyhedra to track the erosion of $\symb{1}$s in a monotone eroder is already present in Toom's original paper.  See the paper of Bramson and Gray~\cite{Berger1966} for a somewhat clearer formulation.  The family of $d$-dimensional polyhedra can be defined by $d+1$ unit vectors $\vec{u}_0,\vec{u}_1,\ldots, \vec{u}_d\in\RR^d$ that are normal vectors of the faces of each polyhedron.  Each polyhedron can thus be specified by $d+1$ points $q_0,q_1,\ldots,q_d\in\RR^d$, where $q_i$ is a point on the $i$'th face of the polyhedron.  We denote such a polyhedron by $R(q_0,q_1,\ldots,q_d)$.  The erosion property of $D$ is now described as follows:
\begin{equation}
    \Delta(\unifO,z)\subseteq R(q_0,q_1,\ldots,q_d)
    \qquad\Longrightarrow\qquad
    \Delta(\unifO,Dz)\subseteq R(q_0-\alpha_0\vec{u}_0,q_1-\alpha_1\vec{u}_1,\ldots,q_d-\alpha_d\vec{u}_d)
\end{equation}
where $\alpha_i$ is the speed of erosion along $-\vec{u}_i$.  We allow some $\alpha_i$'s to be arbitrary real numbers (positive, negative, zero), but the erosion property requires that $R(q_0-s\alpha_0\vec{u}_0,q_1-s\alpha_1\vec{u}_1,\ldots,q_d-s\alpha_d\vec{u}_d)=\varnothing$ for some $s\geq 0$.  The points $q_0,q_1,\ldots,q_d$ along with a starting time $t$, also define a $(d+1)$-dimensional polyhedron
\begin{equation}
    L(q_0,q_1,\ldots,q_d; t) \isdef
        \bigcup_{s\geq 0} R(q_0-s\alpha_0\vec{u}_0,q_1-s\alpha_1\vec{u}_1,\ldots,q_d-s\alpha_d\vec{u}_d)\times\{t+s\} \;,
\end{equation}
which is a pyramid with base $R(q_0,q_1,\ldots,q_d)$ at time~$t$.

Let $E$ denote the set of space-time positions at which an error occurs.  We call this the \emph{noise}.  It is instructive to also consider non-random noise.  Let $Z$ be a trajectory with initial configuration $z$ and noise~$E$.
The \emph{disagreement set} of $Z$ is the set $\Delta(\unifO,Z)=\{(k,t): Z_{k,t}=\symb{1}\}$ of space-time positions at which $Z$ deviates from the all-$\symb{0}$ trajectory.

The erosion property ensures that
\begin{itemize}
    \item If $\Delta(\unifO, z)$ is included in a polyhedron $R(q_0,q_1,\ldots,q_d)$ and the noise~$E$ is contained in the pyramid $L(q_0,q_1,\ldots,q_d; 0)$, then the disagreement set $\Delta(\unifO,Z)$ also remains contained in the same pyramid $L(q_0,q_1,\ldots,q_d; 0)$.
\end{itemize}
The main lemma of Gács and Reif~\cite[Lemma~5.1]{GR1988} formulates the idea that
\begin{itemize}
    \item If $\Delta(\unifO, z)$ is included in a polyhedron $R(q_0,q_1,\ldots,q_d)$ and the noise~$E$ is ``sufficiently sparse'' outside the pyramid $L(q_0,q_1,\ldots,q_d; 0)$, then the part of the disagreement set ``caused by''~$\Delta(\unifO, z)$ remains contained in a slightly taller pyramid~$L_\delta(q_0,q_1,\ldots,q_d; 0)$, where
    \begin{equation}
        L_\delta(q_0,q_1,\ldots,q_d; t) \isdef
            \bigcup_{s\geq 0} R\big(q_0-s(\alpha_0\pm\delta)\vec{u}_0,q_1-s(\alpha_1\pm\delta)\vec{u}_1,\ldots,q_d-s(\alpha_d\pm\delta)\vec{u}_d\big)\times\{t+s\}
    \end{equation}
    for some small~$\delta>0$, where the signs in front of $\delta$ are chosen so as to make the set larger.
    (That is, the erosion still occurs but at a slightly lower speed.)
    See below for a more precise formulation.
\end{itemize}
For a finite set $A\subseteq\ZZ^d\times\ZZ$, let $L(A)$ and $L_\delta(A)$ denote the smallest pyramids of the above form that contain~$A$.  Clearly, $A\subseteq L(A)\subseteq L_\delta(A)$.

The notion of sparseness, due to Gács, has proven very powerful in studying the stability properties of CA and related models with noise~\cite{Gacs1986,GR1988,Wang1991,Gacs2001,DRS2012,Taati2015,CG2021,FMT2022}, as an alternative to the contour arguments more common in statistical physics.  Here, we use one of the several variants.

An \emph{$(r,R)$-island} of a set $E\subseteq\ZZ^d\times\ZZ$ is a set $A\subseteq E$ whose diameter is at most~$r$ and whose distance from $E\setminus A$ is more than~$R$.  We call the set of sites within distance $R$ from $A$ the \emph{territory} of~$A$.
Sparseness is defined in a hierarchical manner, using an appropriately chosen sequence $(\ell_n)_{n\geq 0}$ and a constant $c>0$.  Consider $E\subseteq\ZZ^d\times\ZZ$.  Let $E_0\isdef E$ and, for $n\geq 0$, recursively construct $E_{n+1}$ from $E_n$ by deleting all its $(\ell_n,c\ell_n)$-islands.  We call the $(\ell_n,c\ell_n)$-islands of $E_n$ the islands of \emph{level~$n$}, and denote by~$\family{C}_n$ the collection of all such islands.  With some abuse of terminology, we call the elements of $\family{C}\isdef\bigcup_{n\geq 0}\family{C}_n$ the \emph{islands} of $E$.  We say that $E$ is \emph{sparse} (with respect to $(\ell_n)_{n\geq 0}$ and $c$) if
\begin{enumerate}[label={(\roman*)}]
    \item $\bigcap_{n\geq 0} E_n=\varnothing$ (i.e., every element of $E$ is eventually part of an island and is deleted).
    \item Each site $k\in\ZZ^d$ belongs to the territory of at most finitely many islands.
\end{enumerate}

Let $E\subseteq\ZZ^d\times\ZZ$ be sparse.  For $s\in\ZZ$, we let $E_{>s}\isdef E\cap(\ZZ^d\times\ZZ_{>s})$.  The \emph{birth time} of an island $A$ of $E_{>s}$ is the time of the base of the pyramid~$L(A)$.  We say that $A$ is \emph{primal} in~$E_{>s}$ if its birth time is earliest among all the islands of~$E_{>s}$.
The following is a reformulation of the main lemma of Gács and Reif~\cite[Lemma~5.1]{GR1988} and holds for every $\delta>0$ and $c>0$ and appropriate choices of $(\ell_n)_{n\geq 0}$:
\begin{lemma}[Effect of a primal island of noise in a sparse background]
    Let $A$ be a primal island in~$E_{>0}$.  Let $X$ and $Y$ be perturbed trajectories of~$D$ with common initial configuration~$\unifO$, and with noise sets~$E_{>0}$ and $E_{>0}\setminus A$, respectively.
    Then, $\Delta(X,Y)\subseteq L_\delta(A)$.
\end{lemma}
By successively deleting the primal islands of~$E_{>0}$, we find that, if $X$ is a trajectory of~$D$ with initial configuration~$\unifO$ and noise set~$E_{>0}$, then $\Delta(\unifO,X)\subseteq\bigcup L_{\delta}(A)$, where the union is over all the islands $A$ of~$E_{>0}$.

Now, for $s\in\ZZ$, let $Z^{(s)}$ be a space-time configuration such that $Z^{(s)}_{a,b,k,t}=\symb{0}$ for~$t\leq s$ and $Z^{(s)}$ forms a perturbed trajectory of~$D$ with noise set~$E_{>s}$ for $t\geq s$.  Using a monotonicity argument, one can see that, as $s\to-\infty$, $Z^{(s)}$ converges almost surely to a space-time configuration~$Z$ that is the smallest (with respect to~$\preceq$) bi-infinite perturbed trajectory of~$D$ with noise set~$E$.  The above lemma implies that
\begin{equation}
    \Delta(\unifO,Z)\subseteq\bigcup_{A\in\family{C}} L_{\delta}(A) \;,
\end{equation}
where $\family{C}$ is the family of islands of~$E$.

Theorem~\ref{thm:toom:sea-island} now follows from the following lemmas, which hold for every $\delta>0$, $r\in\NN$, and $c>0$, and suitable choices of $(\ell_n)_{n\geq 0}$:
\begin{lemma}[Sparseness of Bernoulli random sets; see~\cite{DRS2012}]
    For all sufficiently small~$\varepsilon>0$, an $\varepsilon$-Bernoulli random set $E\subseteq\ZZ^d\times\ZZ$ (i.e., the noise of a uniform $\varepsilon$-perturbed trajectory) is almost surely sparse.
    Furthermore, as $\varepsilon\downarrow 0$, the probability that a given position $(k,t)$ belongs to the territory of at least one island goes to~$0$.
\end{lemma}
\begin{lemma}[Sea-island picture of pyramids around a sparse set]
    Let $E\subseteq\ZZ^d\times\ZZ$ be sparse with island decomposition~$\family{C}$.
    Then, $\bigcup_{A\in\family{C}} L_{\delta}(A)$ has the range-$r$ sea-island picture.
\end{lemma}

\end{document}